\documentclass[5p,twocolumn]{elsarticle}

\usepackage{graphicx}
\usepackage{dcolumn}
\usepackage{pifont}
\usepackage{bm}
\usepackage{multirow}
\usepackage{amsmath}
\usepackage{float}
\usepackage{txfonts}
\usepackage[version=3]{mhchem}
\usepackage[usenames,dvipsnames]{xcolor}

\definecolor{cream}{RGB}{222,217,201}
\definecolor{red}{RGB}{225,0,0}

\journal{Mater. Chem. Phys.}

\begin{document}
\title{Formation and characterization of ceramic coating from alumino silicate mineral powders in the matrix of cement composite on the concrete wall}

\author[kimuniv]{Chol-Jun Yu\corref{cor}}
\ead{cj.yu@ryongnamsan.edu.kp}
\author[kumsu]{Byong-Hyok Ri}
\author[kimuniv]{Chung-Hyok Kim}
\author[kimuniv]{Un-Song Hwang}
\author[kimuniv]{Kum-Chol Ri}
\author[kimuniv-anal]{Chang-Jin Song}
\author[kimuniv-anal]{Un-Chol Kim}

\cortext[cor]{Corresponding author}

\address[kimuniv]{Faculty of Materials Science, Kim Il Sung University, Ryongnam-dong, Taesong District, Pyongyang, Democratic People's Republic of Korea}
\address[kumsu]{Changhun-Kumsu Agency for New Technology Exchange, Mangyongdae District, Pyongyang, Democratic People's Republic of Korea}
\address[kimuniv-anal]{Institute of Analysis, Kim Il Sung University, Ryongnam-dong, Taesong District, Pyongyang, Democratic People's Republic of Korea}

\begin{abstract}
Enhancement of thermal performance of concrete wall is nowadays of great importance in reducing the operational energy demand of buildings.
We developed a new kind of inorganic coating material based on \ce{SiO2}-\ce{Al2O3}-rich minerals and Portland cement (PC) powder.
The finely pulverized mineral powder with the particle size distribution (PSD) of 0.4-40 $\mu$m was mixed with the vehicle solvent containing some agents, cement powder with PSD of 2-100 $\mu$m, and water in the certain weight ratio, producing the colloid solution.
After application within 2 hours to the plaster layer of concrete wall and sufficient long hardening period of over three months, the coating layer of 0.6-1.0 mm thickness was observed to become a densified ceramic.
Powder X-ray diffraction (XRD) experiments were performed to identify the crystalline components of minerals, cement and ceramic coating powders.
Three- and two-dimensional surface morphologies and chemical compositions of coating material were obtained with the optical interferometer and scanning electron microscope (SEM) equipped with an energy dispersive X-ray analyzer (EDX).
These XRD and SEM/EDX analyses demonstrated obviously that the coating layer is mainly composed of the calcium-silicate-hydrate (C-S-H) and the calcium-aluminate-hydrate (C-A-H) ceramics with the relatively small number of closed pores (10\% porosity) compared with the cement mortar and concrete layers.
Two-step hydrations of cement and subsequently \ce{SiO2}-\ce{Al2O3} promoted by the alkali product \ce{Ca(OH)2} were proposed as the main mechanism of ceramic formation.
\end{abstract}

\begin{keyword}
Coating \sep Calcium-silicate-hydrate \sep Chemically bonded ceramics \sep Mineral composite
\end{keyword}

\maketitle

\section{Introduction}
Green buildings with zero-energy and zero-carbon are nowadays of great importance in mitigation of global warming and climate change due to large-scale energy consumption and severe environmental impact of building industry~\cite{Ortiz17eb}.
To promote the green building construction with concrete, one should pay primary attention to a decrease in the operational energy since it accounts for almost 80\% of the energy demand of buildings~\cite{Koezjakov18eb}.
In this context, extensive research has been performed to find a way of enhancing thermal insulation performance of the concrete wall to reduce a heating/cooling load of buildings.

It has been reported that thermal insulation of the concrete wall can be improved by altering microstructure and composition of concrete itself~\cite{Sabnis12} or attaching heat insulation boards to its external surface.
Here, the heat insulation boards with a thickness of several centimeters are made from high thermal insulation materials such as expanded perlite~\cite{Jiesheng16eb}, polystyrene foam~\cite{Bandala15ijes} and silica aerogel~\cite{Gao16be}.
They are commonly characteristic of low thermal conductivity below 0.1 W/(m$\cdot$K), so that the heat transfer by a conductive way can be effectively obstructed.
However, these materials have some crucial problems such as high production cost, complicated construction procedure and relatively low durability.
Moreover, some of these materials contain organic agents that may readily decay and be volatile, resulting in considerable harm to the residents' health.

Another way of enhancing thermal performance of the building wall is to paint it with cool materials or coatings~\cite{Chenn16ci,Neto16cbm,Cekon14eb}, which can remarkably reduce the construction cost due to relatively low cost of raw materials and simple procedure.
Coatings with specially designed thermal reflectance and infrared emittance are recognized to effectively prevent the heat transfer from outside to inside of building and vice versa~\cite{Karlessi09se,ZZhang15eb}.
In particular, inorganic mineral-based coatings have been found to exhibit outstanding thermal insulation performance, together with other potential benefits such as environmental benignity and healthy profit~\cite{Kolokotsa12se,Barrera14eb}.
For instance, coatings composed of dolomite or limestone marble powder and glass beads were shown to enhance the infrared emittance of the concrete wall and reduce the solar reflectance, resulting in the significant reduction of cooling load by almost 20\% on annual basis~\cite{Gobakis15eb}.
Nevertheless, there remain several issues for the inorganic coatings to the concrete wall, such as degradation of their performance by weathering and corrosion~\cite{Santamouris08se} and further improvement of thermal insulation performance.

These motivate us to develop a new kind of inorganic composite coating materials with long-term stability and affordable production cost~\cite{Ri18wo}.
These new coatings are in the form of colloid solution, made by mixing various \ce{SiO2}-\ce{Al2O3}-rich mineral powders as fillers, Portland cement powder as matrix and aqueous solution with some additive agents.
Both the external and internal surfaces of the concrete wall were brushed or sprayed with this colloid solution to form a coating layer with a thickness of 0.6-1.0 mm on the wall.
After three months of hardening and growing, the coating was found to become ceramic, exhibiting multiple positive functions such as thermal insulation, water proofing and fire proofing.
Here, formation of the ceramic coating on the concrete wall is the key to get some understanding of such various functions.
Wan et al.~\cite{YWan14cbm} showed that the coating with characteristics of sand texture and imitation ceramic had good adhesive, corrosion, acid and alkali resistance, increased degradation temperature up tp 410 $^\circ$C, and low thermal conductivity of 0.055 W/(m$\cdot$K).
How can the ceramic coating be formed without calcination or pressing?
Zhu et al.~\cite{Zu15mcp} investigated the calcium aluminate cement with micro-sized spinels, indentifying the first stage of cement hydration.
Zinatloo-Ajabshir et al.~\cite{Ajabshir17ijhe,Ajabshir19jem,Ajabshir18jac,Ajabshir17spt,Ajabshir17jec, Salehi17jre,Razi17jml,Beshkar17po} emphasized the substantial impact of particle shape and size on the properties of nanostructured inorganic oxide materials.

Although the hydration and hardening processes of cement has been widely studied and their importance to the properties of the concrete is well understood, the research on the mineral-cement based ceramic formation has not been performed to the best of our knowledge.
In this work, based on the preparation of ceramic coatings on the concrete wall, the experimental method would be applied to identify their crystalline components, chemical compositions and surface morphology, using X-ray diffraction (XRD) and scanning electron microscope (SEM) combined with energy dispersive X-ray (EDX) analysis.
The purpose of this research is to describe the clear stage of cement hydration in the existence of alumino silicate minerals, which would be helpful for understanding of ceramic formation and the application of ceramic coating in building industry.

\section{Materials and methods}
\subsection{Raw mineral materials}
In this work, the raw materials of coating to the concrete wall included several kinds of \ce{SiO2}-\ce{Al2O3}-rich minerals such as silicate, silica sand (quartz), mica, nickel-zinc ferrite and swelling clay (or bentonite), and additive minerals such as graphite and ultraanthracite ash.
In particular, we made use of ``Kumgang'' medical stone (quartz-adamellite), which is a kind of composite mineral that contains plagioclase, K-feldspar, biotite and quartz~\cite{Wang12ogr}.
All the raw mineral materials were mined in several regions of the Korean peninsula, and processed through crushing, sieving and dressing.
The processed raw materials were ball milled in dry state for a sufficient time of 3 hours to produce ultrafine powders.
In order to sift particles with desirable sizes, a fan blower was usd to blow the mineral powder, and the particles fallen at the certain distance range were collected.

\subsection{Particle size distribution analysis}
The particle size distribution (PSD) of mineral and cement powders were analyzed using a laser scattering method as implemented in BT-9300H Laser Particle Size Analyzer (Dandong Bettersize Instruments Ltd., China), which has a capability to measure the size range of 0.1$\sim$340 $\mu$m with a repeatability of $\leq1$\% and a test time of 1$\sim$3 minutes.
The mineral powder of 4 g mass was put into the test container and the distilled water of 250 ml was added to the circulating tank.
Then, ultrasonic and circulating were turned on and runned for 3 minutes.
For the case of cement powder, ethyl alcohol was used instead of water to inhibit the chemical reaction.

\subsection{Preparation of coating paint and samples}
Upon the preparation of mineral powders and cement powder, a vehicle was prepared as aqueous solution by using several kinds of solvents.
They included triethanolamine as surface active agents (2$\sim$5 wt.\%), diethanolamine lauryl sulphate or albumino-exudate, pulp waste liquor and diatomite as air entrained agents (5$\sim$10 wt.\%), hydroxy ethyl cellulose as a viscosifier (1$\sim$2 wt.\%), and tributyl phosphate or butanol as foam elimination agents (1$\sim$2 wt.\%).

The mineral powder was added to this solution in a 1:2 wt. ratio, producing a colloid solution with a dark brown color.
The Portland cement powder was mixed with the water in a 1:2 wt. ratio to form a cement slurry, and then mixed with the mineral colloid solution in a 1:2 wt. ratio to produce a cement-mineral composite slurry.
Either a hand mixing stick or a mixing machine was used for the stirring work.

The concrete wall should be covered well with cement mortar plaster of 20$\sim$25 mm thickness.
After the plaster surface hardened, the cement-mineral composite slurry was broadly sprayed with an air-pump spray gun (or brushed with a hand brush) on the external and internal surfaces over two times within two hours of its preparation.
Thickness of the coating layer should be about 1 mm.
To prevent surface drying and accelerate hardening of the coating layer, water could be sprinkled three times at an interval of 2 hours.
After three months of hydration and hardening, small chips of 15$\times$15$\times$5 mm sizes were taken off from the external and internal surfaces of the coated wall to be used as samples for material characterization.

\subsection{Material characterization}
Three-dimensional (3D) surface morphologies of the coated concrete wall samples were analyzed using the MicroXAM interferometer.
The crystalline components of materials (mineral, cement, and cement-mineral composite) were determined from powder XRD measurements using a Cu K$\alpha$ (1.54051 \AA) radiation source and operating at a speed of 2$^{\circ}$/min from the Smartlab X-ray deffractometer (Rigaku, Japan).
Their chemical compositions were identified by EDX analysis using the ZSX Prismus III$^{+}$ instrument (Rigaku, Japan).
Thermogravimetric analysis (TGA) with differential thermal analysis (DTA) was performed with the instrument of Shimadzu TGA-50H (Japan) at a heating rate of 10$^\circ$C/min from 20 to 1000 $^\circ$C.
SEM images of the surface were characterized using the ZSM-6610A (JEOL, Japan) field-emission gun scanning electron microscope at an accelerating voltage of 20 kV, which integrates the energy dispersive spectrometer (EDS) for elemental analysis or chemical composition of samples.

\section{Results and discussion}
\subsection{PSD analysis}
\begin{figure}[!th]
\centering
\includegraphics[clip=true,scale=0.47]{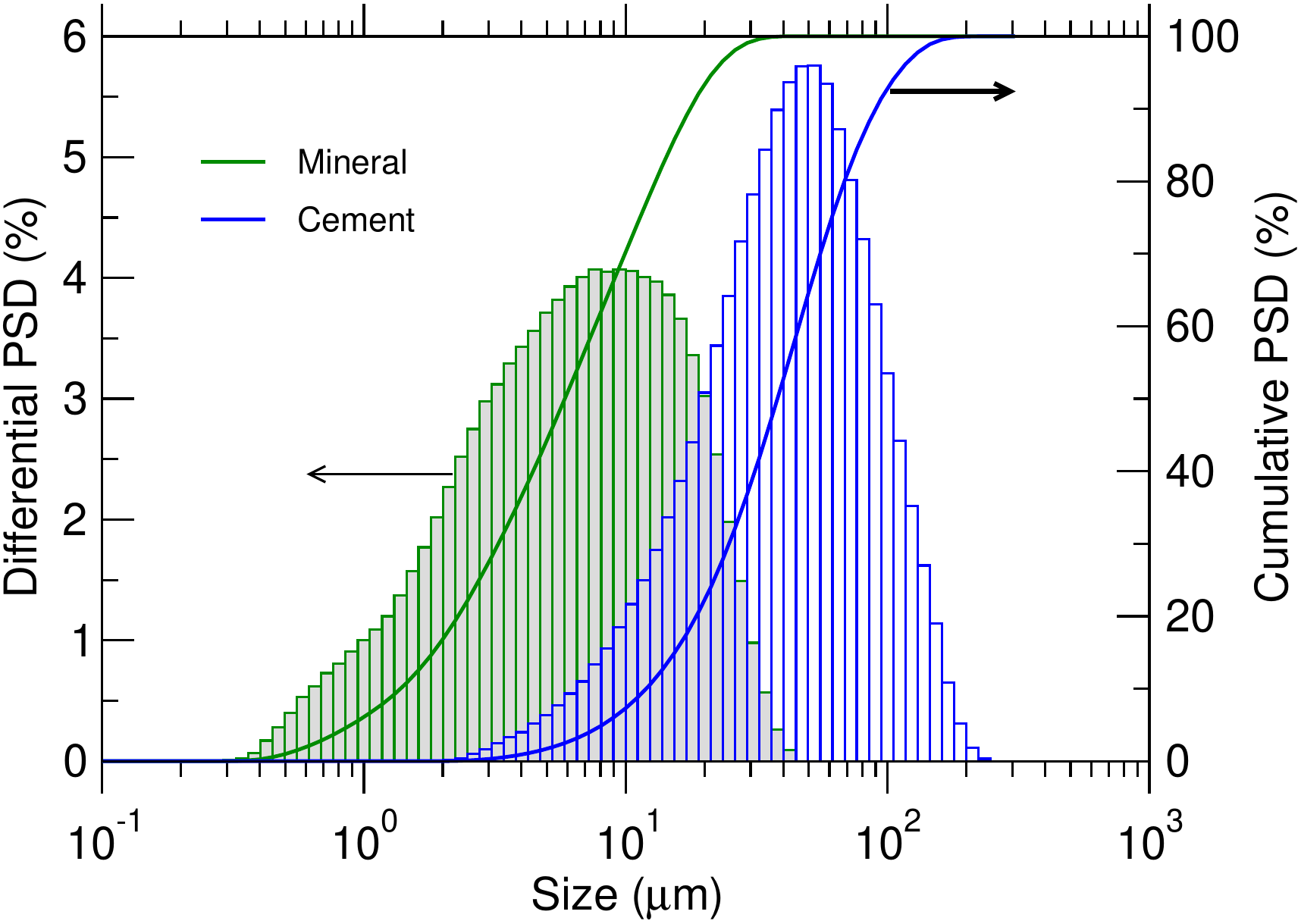}
\caption{\label{fig_psd}Differential and cumulative particle size distributions (PSD) of mineral and cement powders. The particle sizes of mineral powder range from 0.4 to 40 $\mu$m with an average value of 8 $\mu$m, while those of cement powder range from 2 to 110 $\mu$m with an average value of 50 $\mu$m.}
\end{figure}

The powder PSD is decisive for the quality and performance of composite coating material, such as hardening time, ultimate strength, pore size and distribution, mineral filling, coating effect and surface morphology.
Therefore, it is important to effectively control and precisely determine PSDs of the raw materials.
Upon production of the mineral powder using the ball mill method, the distribution range of their particle sizes was obtained by analyzing the scattering pattern of particles generated on the focal plane on Fourier lens when illuminated by a parallel and monochromatic laser light.
Figure~\ref{fig_psd} shows the determined PSDs of the mineral and cement powders.
It can be seen that the particle sizes of mineral powder range from 0.4 to 40 $\mu$m with an average size of 8 $\mu$m, while those of cement powder range from 2 to 100 $\mu$m with an average size of 50 $\mu$m.

In the conventional cement-based composite materials, the Portland cement is a binder or matrix, while the mineral powder might be an admixture or additive, according to the traditional classification.
However, the mineral powder in this work accounts for almost 35\% of composite, indicating that the fillers or microfillers are more appropriate to be called for mineral component rather than additives.
When the cement slurry is formed with water, the exothermic chemical reaction of cement hydration occurs, leading to a formation of hydrate crystallites such as \ce{Ca(OH)2} and \ce{CaO$-$SiO2$-$H2O} (C$-$S$-$H) with low solubilities, which grow slowly or quickly to become large crystals with sizes ranging from 10 $\mu$m to 1.0 mm~\cite{Maruyama15ccr}.
The products of cement hydration, C$-$S$-$H, are called cement gel, which is an aggregation of colloidal material.
As the hydration progresses, the neighboring grains get in contact and a dense gel structure fills available spaces, leaving thin gel pores of 1$\sim$10 nm size in the gel structure and capillary pores of 0.1$\sim$1000 $\mu$m size between the gel layers~\cite{Brandt09}.
Since the sizes of mineral particles in this work were smaller than those of cement powder, the major mineral compounds \ce{SiO2} and \ce{Al2O3} may cause the second hydration reaction with \ce{Ca(OH)2} extracted from the hydration of cement, forming additional hydrate gel to fill the pores and become dense ceramics.

\subsection{Composition analysis}
\begin{figure}[!th]
\begin{center}
\includegraphics[clip=true,scale=0.53]{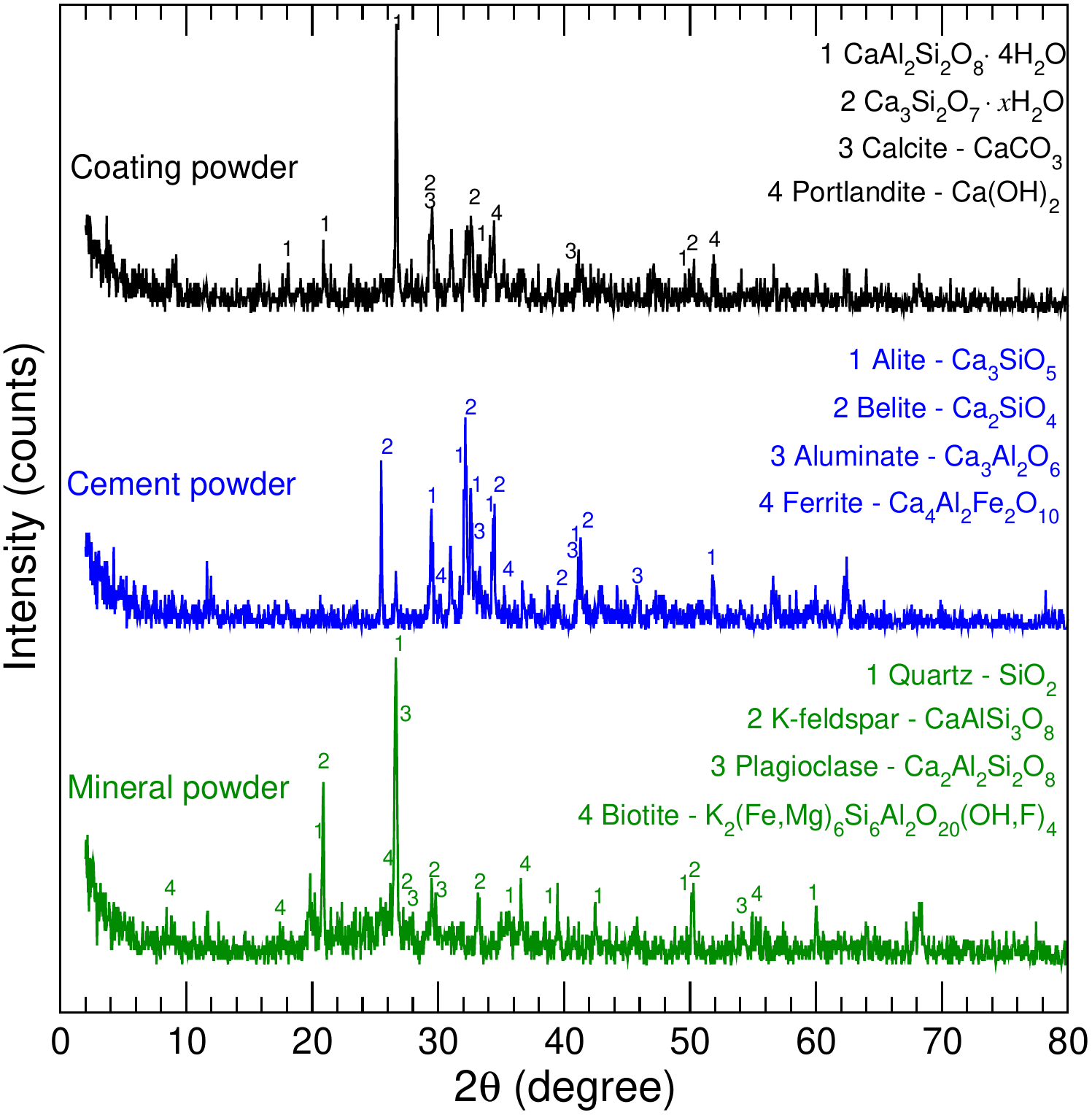}
\caption{\label{fig_xrd}XRD patterns of mineral, cement and coating ceramics powders. XRD peaks correspond to crystals of quartz-adamellite for mineral, typical cement crystals for cement, and some hydrates for coating.}
\end{center}
\end{figure}

Figure~\ref{fig_xrd} shows XRD patterns of the mineral, cement, and coated sample powders (the sample surface was scratched to obtain coating powder).
It was known that identifying all kinds of crystals included in mineral powders by XRD is difficult.
For the case of mineral powder, some typical XRD peaks were found to correspond to the components of Kumgang medical stone, such as quartz (\ce{SiO2}), K-feldspar (\ce{CaAlSi3O8}), plagioclase (\ce{Ca2Al2Si2O8}), biotite, etc.
It should be noted that the XRD patterns of individual component minerals indicated other crystals such as hematite, corundum, cuprite, dolomite, calcite, muscovite, kalininite, chlorartinite, and graphite.
For the case of cement powder, one can see typical XRD peaks corresponding to alite (\ce{Ca3SiO5}), belite (\ce{Ca2SiO4}), aluminate (\ce{Ca3Al2O6}), and ferrite (\ce{Ca4Al2Fe2O10}).
By careful analysis of XRD patterns of the composite coating powder, some hydrates including calcium silicate hydrates and calcium alumino silicate hydrates were identified, together with carbonates (calcite and dolomite) and calcium hydroxide (\ce{Ca(OH)2}).

The concentrations of major metal oxides in the same samples, obtained by means of EDX analysis, are shown in Table~\ref{tab_edx}.
The major oxides were found to be \ce{SiO2} (57.2 wt.\%), \ce{Al2O3} (20.5 \%), \ce{Fe2O3} (8.9 \%), \ce{K2O} (3.6 \%), \ce{SO3} (3.3 \%), and CaO (3.0 \%) in the mineral powder, while for the cement powder they were CaO (64.2 \%), \ce{SiO2} (22.9 \%), \ce{Al2O3} (5.3 \%), and \ce{Fe2O3} (3.1 \%) as in general cement.
In the composite coating material, it might be said that somewhat mixing effect was found because the average composition ratios were observed to be \ce{SiO2} (37.3 \%: average 40.1 \%), CaO (35.1 \%: 33.6 \%), \ce{Al2O3} (8.7 \%: 12.9 \%), and \ce{Fe2O3} (7.5 \%: 6.1 \%).
It should be noted that such weight ratio can be slightly variable according to the condition of mixing and production.
\begin{table}[!b]
\small
\caption{\label{tab_edx}Major oxide contents of mineral, cement, and composite powders identified by EDX analysis using the ZSX Prismus III$^{+}$ spectrometer. The components with weight percent under 0.1 \% are ignored. LOI means loss in ignition.}
\begin{tabular}{lc@{\hspace{5pt}}clc@{\hspace{5pt}}c@{\hspace{5pt}}clc}
\hline
\multicolumn{2}{c}{Mineral} & & \multicolumn{3}{c}{Cement} & & \multicolumn{2}{c}{Composite} \\
\cline{1-2} \cline{4-6} \cline{8-9}
Oxide & wt.\% & & Oxide & wt.\% & Ref.~\cite{Parisatto15jms} & & Oxide & wt.\% \\
\hline
\ce{SiO2}  & 57.21 & & \ce{CaO}   & 64.17 & 64.65 & & \ce{SiO2}  & 37.31 \\
\ce{Al2O3} & 20.49 & & \ce{SiO2}  & 21.92 & 21.01 & & \ce{CaO}   & 35.10 \\
\ce{Fe2O3} & 8.94  & & \ce{Al2O3} & 5.25  & 4.75  & & \ce{Al2O3} & 8.66 \\
\ce{K2O}   & 3.62  & & \ce{Fe2O3} & 3.14  & 2.37  & & \ce{Fe2O3} & 7.53 \\
\ce{SO3}   & 3.26  & & \ce{SO3}   & 1.85  & -     & & \ce{SO3}   & 5.19 \\
\ce{CaO}   & 2.97  & & \ce{MgO}   & 1.42  & 1.93  & & \ce{K2O}   & 2.49 \\
\ce{MgO}   & 1.51  & & \ce{K2O}   & 0.68  & 0.79  & & \ce{MgO}   & 2.22 \\
\ce{TiO2}  & 1.03  & & \ce{Na2O}  & 0.40  & 0.30  & & \ce{TiO2}  & 0.77 \\
\ce{Na2O}  & 0.66  & & \ce{TiO2}  & 0.13  & 0.13  & & \ce{Na2O}  & 0.36 \\
LOI        & 0.31  & & LOI        & 1.04  & 1.66  & & LOI        & 0.37 \\
\hline
\end{tabular}
\end{table}

In accordance with the composition analysis by XRD and EDX, we can assume the two-step hydrations of cement and further mineral powders.
In the first hydration, the cement compounds, 3\ce{CaO$\cdot$SiO2} (\ce{C3S}: alite), 2\ce{CaO$\cdot$SiO2} (\ce{C2S}: belite) and CaO, are anhydrate compounds with a high reactivity, and thus, when reacting with water, they form the hydrates and calcium hydroxide as follows,
\begin{flalign}
2\ce{Ca3SiO5}+6\ce{H2O} & \rightarrow \ce{Ca3Si2O7\cdot}3\ce{H2O}+3\ce{Ca(OH)2} \label{eq_hydcem1} \\
2\ce{Ca2SiO4}+4\ce{H2O} & \rightarrow \ce{Ca3Si2O7\cdot}3\ce{H2O}+\ce{Ca(OH)2} \label{eq_hydcem2} \\
\ce{CaO}+\ce{H2O} & \rightarrow \ce{Ca(OH)2} \label{eq_hydcem3}
\end{flalign}
where 3\ce{CaO$\cdot$}2\ce{SiO2$\cdot$}3\ce{H2O} (calcium silicate hydrate: C$-$S$-$H) is in the gel state (cement gel) and \ce{Ca(OH)2} is extracted into the capillary pores. As mentioned above, the C$-$S$-$H gel crystallizes into different phases coexisted with etringite and other minor phases, filling the pores initially occupied by water and air during the hydration time~\cite{Parisatto15jms}.

In the second hydration, the main oxides \ce{SiO2} and \ce{Al2O3} of mineral particles with sizes of under 8 $\mu$m have not water-reactivity originally, but in the presence of alkali extract \ce{Ca(OH)2} they can react with water of the pores as follows,
\begin{flalign}
\label{eq_hydmin}
\ce{SiO2}+x\ce{Ca(OH)2}+m\ce{H2O} & \rightarrow x\ce{CaO\cdot}\ce{SiO2\cdot}n\ce{H2O} \\
\ce{Al2O3}+y\ce{Ca(OH)2}+m\ce{H2O} & \rightarrow y\ce{CaO\cdot}\ce{Al2O3\cdot}n\ce{H2O}
\end{flalign}
These additional C$-$S$-$H and C$-$A$-$H gels can fill the capillary pores, resulting in a densification of cement paste and finally formation of cement-mineral based ceramic.
Meanwhile, some \ce{Ca(OH)2} extracts on the surface can react with \ce{CO2} in air on condition of water existence,
\begin{equation}
\label{eq_caoh}
\ce{Ca(OH)2}+\ce{CO2}+n\ce{H2O} \rightarrow \ce{CaCO3}+(n+1)\ce{H2O}
\end{equation}
where the insoluble product \ce{CaCO3} can be placed on the surface, covering the input of pores.

\begin{figure}[!th]
\centering
\includegraphics[clip=true,scale=0.14]{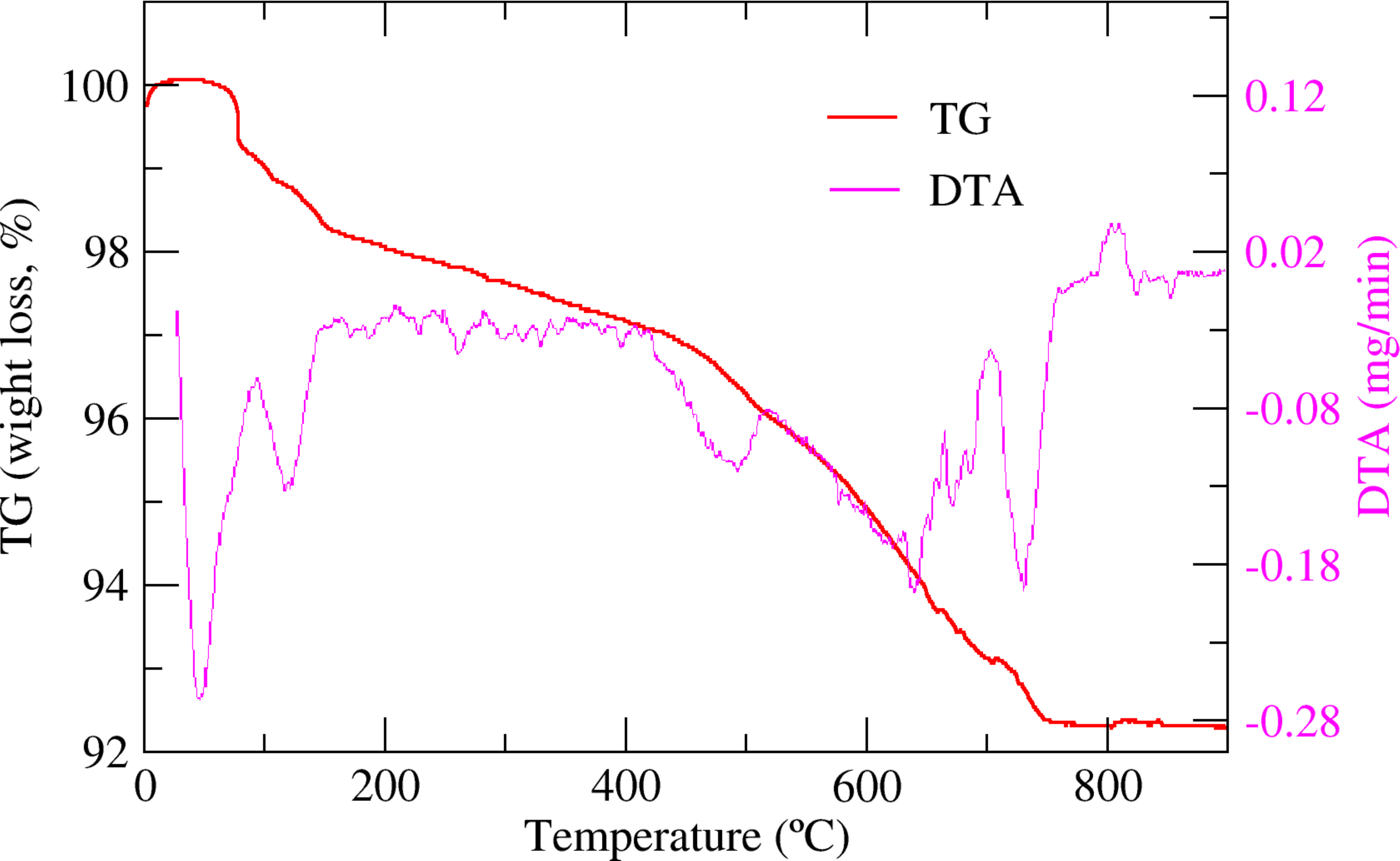}
\caption{\label{fig_tga}Thermogravimetric (TG) and differential thermal analysis (DTA) of coating sample powder.}
\end{figure}
The thermal analysis of the coating sample powder is illustrated in Fig.~\ref{fig_tga}.
The peaks are observed at 78 and 140 $^\circ$C, which can be thought to correspond to the first and second hydrations.
The next peak is found at high temperature of 501 $^\circ$C, indicating the ceramic coating is quite thermally stable and could resist higher temperature in the process of storage or usage.

We measured the porosity of the ceramic coating, mortar layer, and concrete, by applying the following formula
\begin{equation}
P=\left(1-\frac{\rho_0}{\rho_s}\right)\times 100\%
\end{equation}
where $\rho_0$ is the apparent density of the material, calculated from the volume and mass of the sample, and $\rho_s$ is the real density of the material without pore, measured by pulverizing the sample, putting it into the measuring flask with water, and boiling and cooling the solution.
Table~\ref{tab_por} present the measured values.
It is worth noting that the real density of ceramic coating (2.268 g/cm$^3$) is close to the values of typical C$-$S$-$H crystalline solid phases such as jennite \ce{Ca9Si6O18(OH)6\cdot}8\ce{H2O} (2.325 g/cm$^3$)~\cite{Bonaccorsi04ccr} and tobermorite~\ce{Ca5Si6O16(OH)2\cdot}7\ce{H2O} (2.23 g/cm$^3$)~\cite{Bonaccorsi05jacs}, indicating the conversion of mineral powder into cement-based ceramics.
The porisity of ceramics coating was estimated to be $\sim$10\%, being much smaller than those of cement mortar (23.2\%) and cement concrete (35.8\%).
\begin{table}[!th]
\small
\caption{\label{tab_por}Measured apparent density ($\rho_0$), real density ($\rho_s$) and porosity ($P$) of materials.}
\begin{tabular}{lccc}
\hline
Materials & $\rho_0$ (g/cm$^3$) & $\rho_s$ (g/cm$^3$) & $P$ (\%) \\
\hline
Ceramic coating & 2.040 & 2.268 & 10.05 \\
Cement mortar      & 2.053 & 2.674 & 23.22 \\
Cement concrete    & 2.091 & 3.255 & 35.76 \\
\hline
\end{tabular}
\end{table}

\subsection{Surface morphology analysis}
\begin{figure*}[!th]
\centering
\includegraphics[clip=true,scale=0.25]{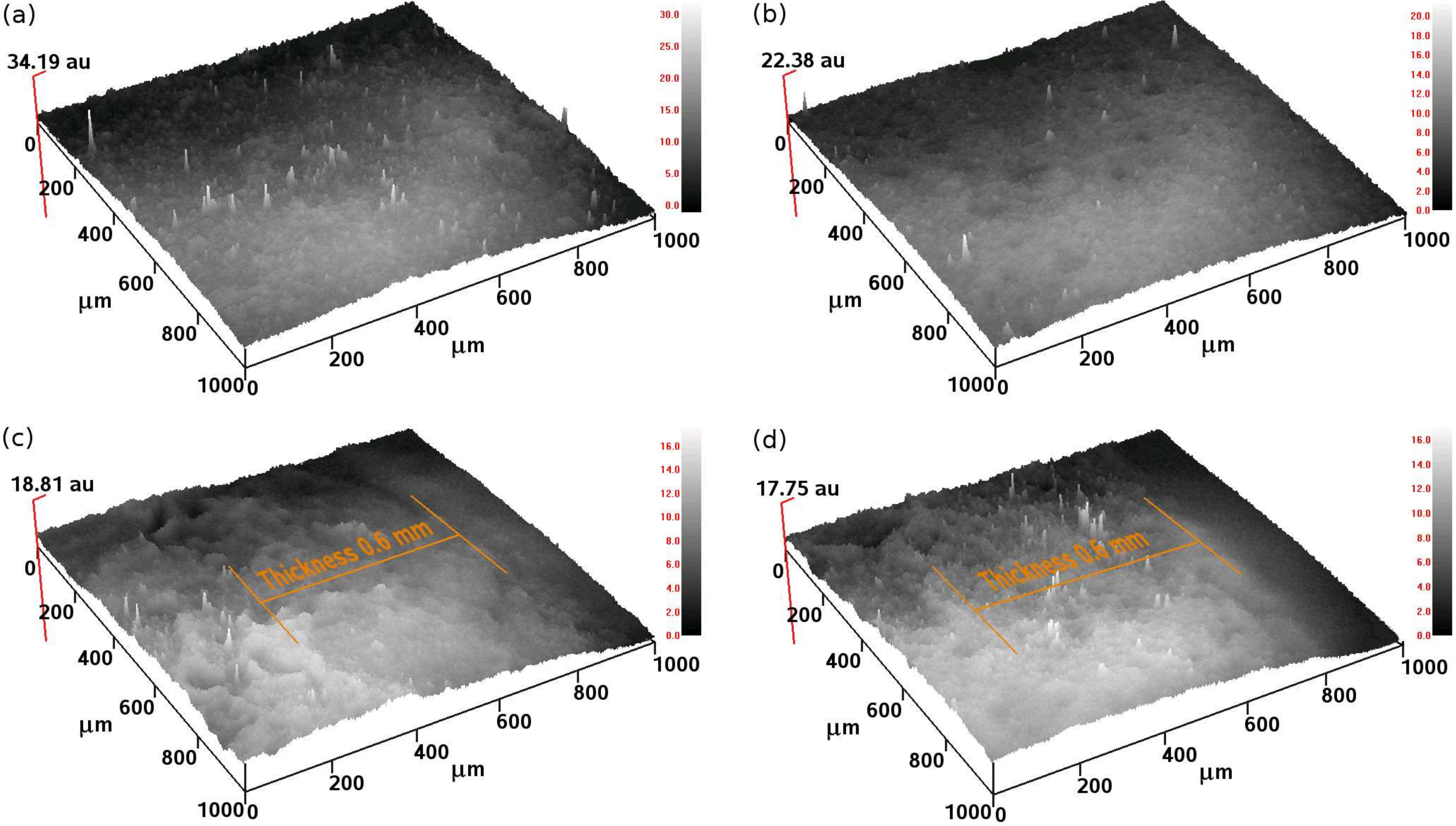}
\caption{\label{fig_morp}3D surface morphologies of external surface of the coated wall in (a) top and (c) side views, and internal surface in (b) top and (d) side views. Thicknesses of ceramic coating layers are measured to be about 0.6 mm. Here, 1 au (atomic unit) equals to 1 Bohr radius ($\sim$0.529 \AA).}
\end{figure*}
In Fig.~\ref{fig_morp} are shown the 3D surface morphologies of outer and inner concrete walls coated by using the cement-mineral based composite ceramics.
The top surfaces of the outer and inner walls were shown to be clearly smoother than their side surfaces, and to be dense without any obvious defects such as pores and cracks.
Moreover, it could be observed that in side views (Fig.~\ref{fig_morp}(c) and (d)) the profiles of the coating layers were flat, whereas those of cement mortar layers were severely wavy, although the wall cuttings were not fully processed to look glossy.
In these figures, the thicknesses of the coating layers were found to be about 0.6 mm for the both walls, and the interfaces between the coating layer and the cement mortar layer could be identified though the borders might not be clear, implying the strong binding of coating particles to cement mortar layers by inward growth of hydrate gel.

\begin{figure}[!th]
\centering
\includegraphics[clip=true,scale=0.15]{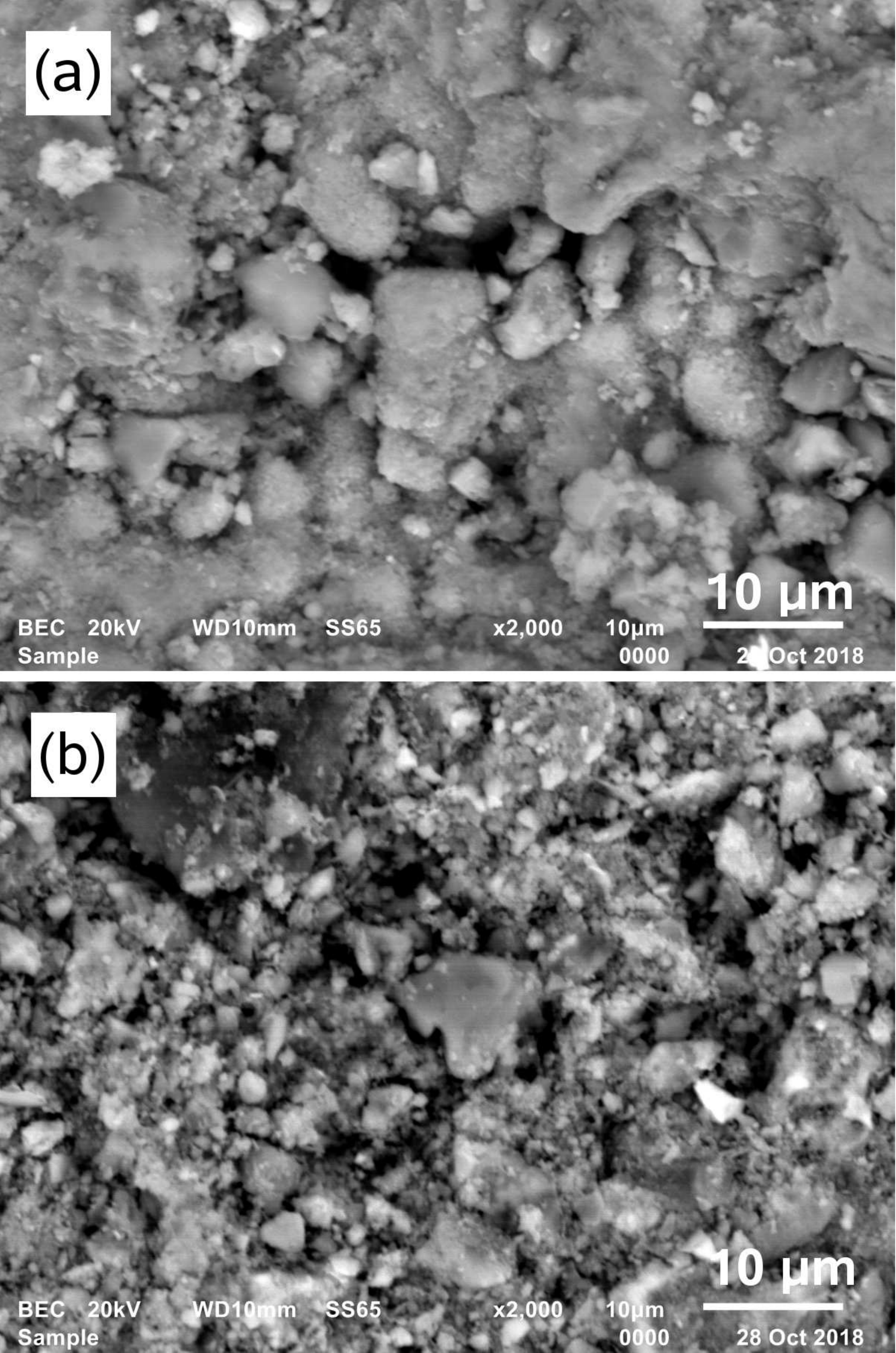}
\caption{\label{fig_bec}SEM images of (a) outer and (b) inner concrete walls coated with cement-mineral based composite ceramics, obtained by using back scattered electrons (BSE).}
\end{figure}
The surface morphologies with a high resolution down to about 1 $\mu$m were obtained with the SEM analysis.
First, we obtained SEM images by using back scattered electrons (BSE), as shown in Fig.~\ref{fig_bec}.
We note that the surfaces were carefully polished and covered with thin gold film to get clear images.
The BSE images reveal that the cement-mineral based ceramic coating contains inhomogeneous grains with a wide spread of sizes from 10 nm to 10 $\mu$m.
It can be seen that the grain sizes of the outer surface are clearly larger than those of the inner wall, indicating further growth of hydrates possibly due to more active contact with moisture in air.
As observed in 3D morphologies, moreover, we can see the continuous matrix regions with an area of over 10 $\mu$m$^2$ without any pore in both the surfaces, especially in the outer wall surface, which can be thought as the result of the aforementioned two-step hydrations.

\begin{figure*}[!th]
\centering
\includegraphics[clip=true,scale=0.55]{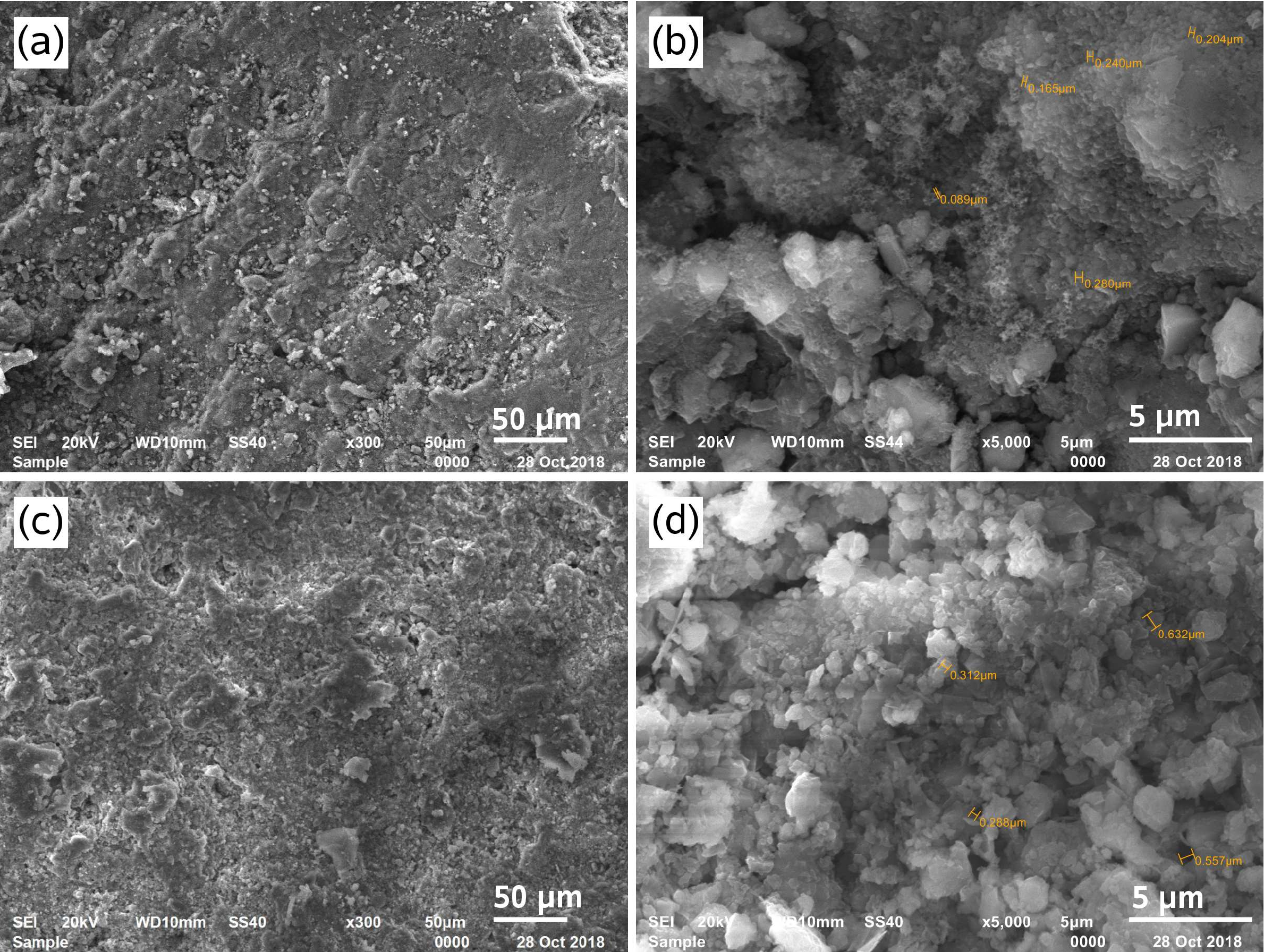}
\caption{\label{fig_sem}SEM images of outer (a, b) and inner (c, d) concrete walls coated with composite ceramics, obtained by using secondary electrons (SE). Sizes of typical grains and pores are marked at higher resolution images (b, d).}
\end{figure*}
Figure~\ref{fig_sem} shows the SEM micrographs (at different amplifications) of the same samples, obtained by using secondary electrons (SE).
From Figs.~\ref{fig_sem}(a) and (c), it is found that though somewhat rough, the coating surfaces have dense structures with no signs of macro-pores and micro-cracks or delaminations.
The rough microstructures, consisting in cloud-like terraces with about dozens of square micrometers, contain nano-sized merged granules, which might be probably induced by the growth of silicate hydrates.
As can be seen in Fig.~\ref{fig_sem}(b) and (d), the coating layer is composed of a lot of tiny particles with sizes of under hundreds of nanometers, being much smaller than the particle size distributions of mineral and cement particles.
Some lumpy particles with sizes of several micrometers were seen to be linked together by the tiny particles or to be attached strongly to the matrix.
It reveals that though some grooves or craters with sizes of micrometers on the surface, they are all closed by below matrix or particles and moreover there are no distinct capillary pores penetrating into the inside.

\begin{figure*}[!th]
\centering
\includegraphics[clip=true,scale=0.22]{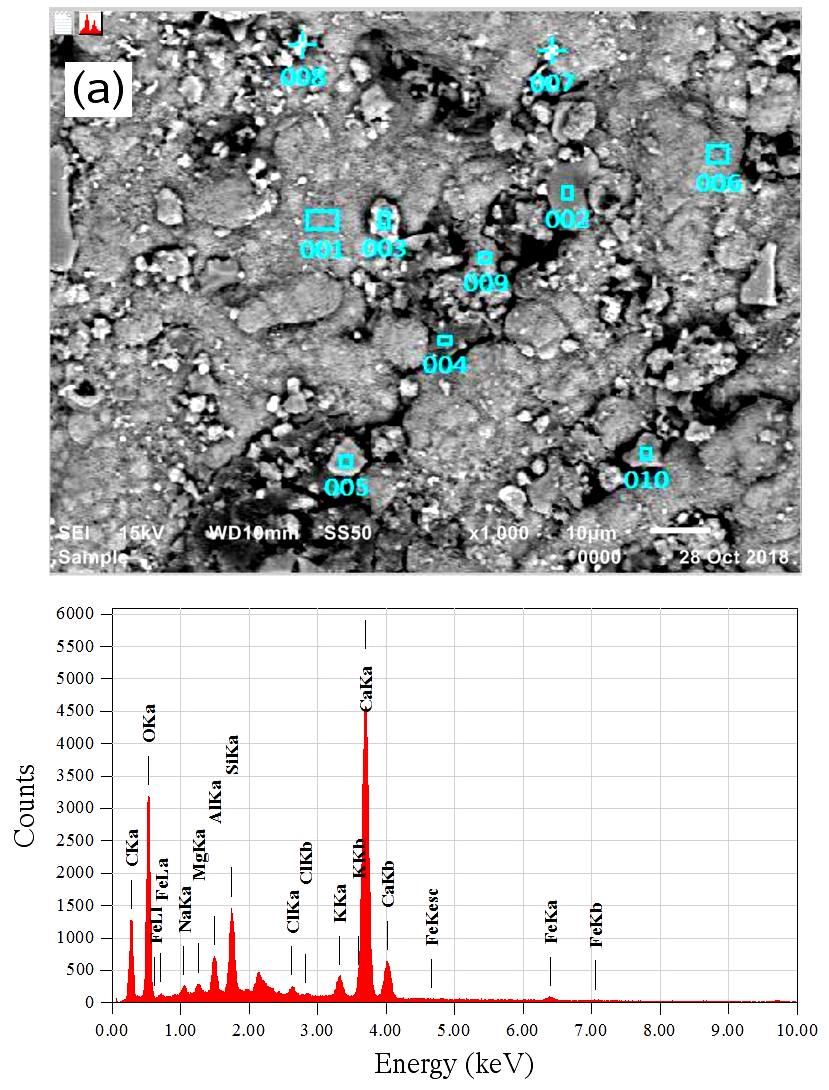}
\includegraphics[clip=true,scale=0.22]{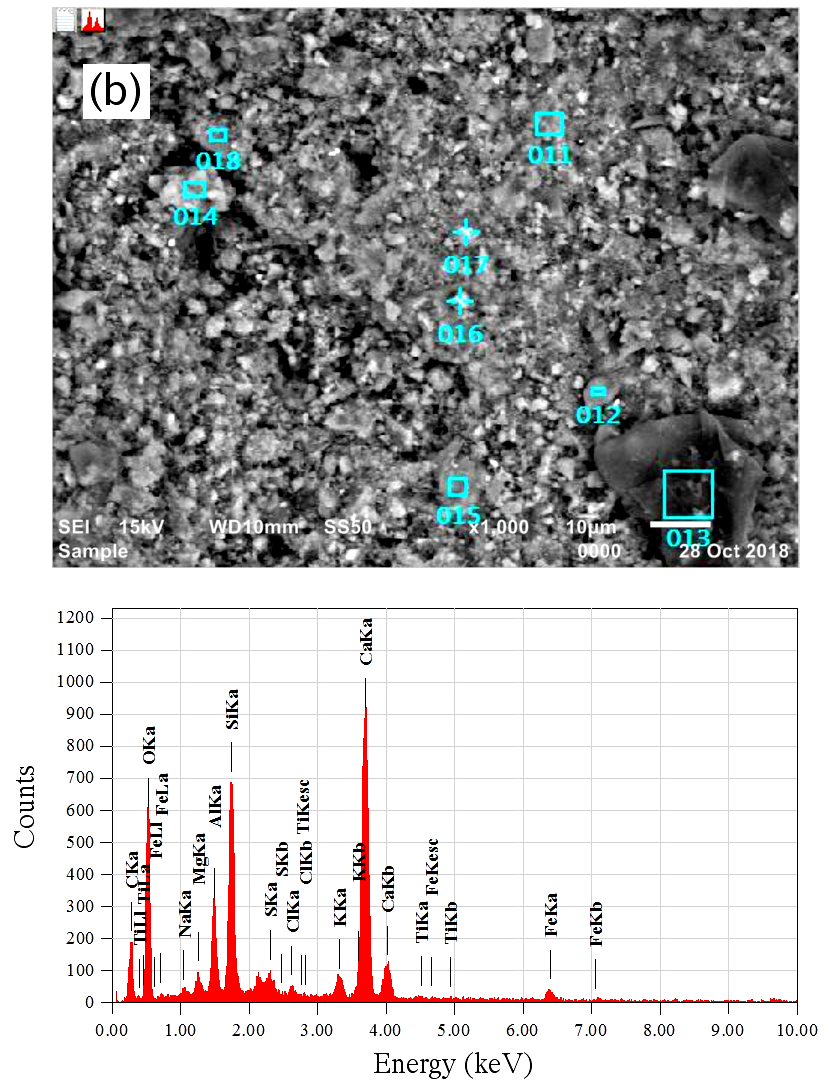}
\caption{\label{fig_eds}BSE-SEM micrographs and typical EDS spectrum of (a) outer and (b) inner concrete walls coated with composite ceramic. Identifying numbers for EDS analysis are denoted on the images.}
\end{figure*}

In order to identify the compositional elements, we performed EDS analysis of the same samples.
In Fig.~\ref{fig_eds} are shown the BSE-SEM images of the outer and inner concrete walls to indicate the measuring points with identifying numbers, and typical EDS spectrum.
Table~\ref{tab_eds} lists the EDS analysis results for compositional elements (in wt.\%) of grains corresponding to these numbers.
On the outer wall surface shown in Fig.~\ref{fig_eds}(a), the bright points with white color (No. 008, 007) were found to contain the highest contents of Ca element (80, 49 \%) and the lowest content of Si (1.5, 1.6 \%), indicating the \ce{CaCO3} crystals.
The grey-colored points such as No. 002 and 010, which occupy the major part of surface, were analyzed to have balanced contents of Ca (18, 18 \%), Si (17, 22 \%), Al (12, 7 \%), and O (22, 25 \%) elements, implying the C$-$A$-$H or C$-$S$-$H crystallized phases.
At the darkest point with No. 005, the Ca content was found to have the lowest value of about 5 \% while the C content to have the highest value of 37 \%, which might be originated from the graphite included in the mineral powder.
In accordance with the EDX analysis (Table~\ref{tab_edx}) that contains the overall composition on the surface, the other compositional elements such as Fe, K, Mg, Na, and S, which could be existed in oxide states, were found not to be negligible.

As shown in Fig.~\ref{fig_eds}(b), the inner concrete wall surface seems to contain few grains of calcium carbonate \ce{CaCO3} because of no white points and no high content of Ca.
Instead, many points with grey color including No. 011, 012, 015 and 016 correspond to the calcium silicate or aluminate hydrate particles due to their balanced contents of Ca, Si, Al, and O elements.
Exceptionally, a distinguished large particle with dark grey color was found at No. 013 point, which contains the highest C element while the lowest Ca element, indicating the intact mineral particle that might not react with water.
It is worth noting that in contrast to the outer wall, the inner surface exhibits distinct portion of Ti element and thus nano \ce{TiO2} particles, which can allow other function such as antibiosis to the wall.

\begin{table*}[!th]
\small
\caption{\label{tab_eds}EDS analysis for compositional elements (wt.\%) of grains corresponding to the identifying numbers denoted in Fig.~\ref{fig_eds}.}
\begin{tabular}{crrrrrrrrrrrr}
\hline
Number & C~~~& O~~~&Na~~&Mg~~&Al~~~&Si~~~&S~~~&Cl~~&K~~~~&Ca~~~&Ti~~&Fe~~ \\
\hline
001 &23.42 &27.59 & 0.22 & 0.41 & 1.15 & 2.39 &      & 0.14 & 1.81 &42.30 && 0.56 \\
002 &16.98 &22.09 & 0.65 & 1.08 &12.31 &17.41 &      & 0.65 & 8.42 &17.52 && 2.90 \\
003 &27.81 &31.36 & 0.67 & 0.55 & 1.13 & 4.20 & 0.05 & 0.40 & 1.56 &31.81 && 0.46 \\
004 &20.66 & 9.77 & 1.45 & 0.86 & 3.58 & 8.78 & 1.67 & 6.14 &11.65 &31.57 && 3.88 \\
005 &37.14 &24.62 & 1.08 & 0.26 & 5.64 &17.03 &      & 0.98 & 7.07 & 5.30 && 0.89 \\
006 &29.74 &33.46 & 0.38 & 0.30 & 1.34 & 1.66 &      & 0.23 & 1.14 &31.56 && 0.19 \\
007 &19.73 &23.35 & 0.60 & 0.14 & 0.77 & 1.55 &      & 1.70 & 3.17 &48.59 && 0.40 \\
008 & 4.66 & 9.62 & 0.30 & 0.11 & 1.06 & 1.48 &      & 0.28 & 1.63 &79.67 && 1.18 \\
009 &22.91 &25.41 & 0.53 & 0.36 & 1.14 & 2.56 & 0.86 & 0.99 & 3.75 &40.80 && 0.69 \\
010 &16.11 &24.69 & 4.85 & 0.36 & 7.47 &21.62 & 0.56 & 0.74 & 4.07 &17.88 && 1.66 \\
\hline
011 &13.06 &53.79 & 0.17 & 1.09 & 2.69 & 5.49 & 0.77 & 0.15 & 0.88 &20.35 & 0.11 & 1.44 \\
012 &16.94 &53.52 & 0.24 & 0.64 & 2.65 & 5.23 & 0.45 & 0.08 & 0.94 &18.29 & 0.08 & 0.93 \\
013 &50.53 &33.32 & 1.69 & 0.39 & 1.10 & 3.06 & 0.49 & 1.80 & 1.20 & 5.82 & 0.07 & 0.53 \\
014 &20.22 &51.46 & 0.28 & 0.56 & 2.76 &15.03 & 0.03 & 0.19 & 1.63 & 6.17 & 0.11 & 1.56 \\
015 & 9.45 &53.51 & 0.25 & 1.22 & 8.28 &13.58 & 0.09 & 0.15 & 3.77 & 7.20 & 0.22 & 2.28 \\
016 &10.49 &58.14 & 0.08 & 0.69 & 3.49 & 6.84 & 0.71 & 0.20 & 1.29 &17.07 & 0.13 & 0.88 \\
017 &11.27 &52.66 & 0.44 & 0.89 & 4.40 & 9.45 & 0.53 & 0.16 & 1.53 &16.84 & 0.08 & 1.73 \\
018 &17.68 &43.12 & 0.22 & 0.37 & 4.31 &23.20 &      & 0.16 & 2.58 & 7.10 & 0.28 & 0.96 \\
\hline
\end{tabular}
\end{table*}

\subsection{Mechanism of ceramic formation}
In the above section, we proposed the two-step hydrations of cement and mineral powders by the action of water as the main mechanism of ceramic formation.
The main products of these hydrations are the C$-$S$-$H and C$-$A$-$H gels, which fill the capillary pores and thus form the dense hardening structure.
The period of hardening time is relatively long over three months, and the complex material processes are occurred in this period.
At the early stage of hydration, the CSH (I) with a distorted wafer phase is created as the metastable phase, and transformed into the coagulated CSH (II) phase with a fibrous shape, exhibiting strength of ceramics.
In fact, as the hydration and hardening of C$-$S$-$H gel progresses, the dehydration-condensation reaction of \ce{Si(OH)4} is repeated as follows,
\begin{equation}
\begin{gathered}
\begin{array}[t]{@{}l @{}c@{}l@{}c@{}l@{}c@{}l@{}c@{}l}
\vspace*{-0.1cm} &\mathsf{l} & & \mathsf{l}& & \mathsf{l}& & \mathsf{l} & \\
\vspace*{-0.1cm}- & \ce{Si} &-\ce{OH-OH}-&\ce{Si}&- \rightarrow - &\ce{Si}&-\ce{O}-&\ce{Si}&-+\ce{H2O}\uparrow \\
&\mathsf{l} & & \mathsf{l}& & \mathsf{l}& & \mathsf{l} & \\ \end{array}
\end{gathered}
\end{equation}
This results in the strong binding between the gel particles through Si$-$O$-$Si bond (silicate group) as bridge with shapes of chain, lamella, and 3D network.
These polymer-like crystallites include a lot of dimers (\ce{Si2O7}) and a relatively small number of fan-shaped trimers (\ce{Si3O10}) and ring-shaped tetramers (\ce{Si4O12}).
It is worth noting that in addition to such Si$-$O$-$Si covalent bond, there exist other chemical bonds such as van der Waals bond, hydrogen bond, and ionic bond induced by inhomogeneity of electronic charge density, which further enhance the strength of densified ceramic structure.

\section{Conclusions}
In this work we investigated the composite coating material based on the cement and mineral powders to identify formation and characterization of ceramic coating on the concrete wall.
Several kinds of \ce{SiO2}-\ce{Al2O3}-rich minerals including Kumgang medical stone were selected and pulverized into fine powder with the controlled particle size distribution ranging from 1$\sim$10 $\mu$m.
Mixing the mineral powder with the vehicle solvent including several kinds of agents produced the colloid solution, which was further mixed with cement slurry to make the composite coating material to the concrete wall.
After applying the coating material to the outer and inner walls and passing three months for hardening, the dense ceramic coating layer was found to form on the surface with a thickness of about 0.6 mm.
We performed the analysis of chemical composition and surface morphology of the ceramic coating material by using XRD and SEM.
It was revealed that the product \ce{Ca(OH)2} from the chemical reaction of cement powder with water could facilitate the hydration of mineral powder, resulting in the second production of the C$-$S$-$H and C$-$A$-$H gels that fill the capillary pores.
We provided the discussion of ceramic formation mechanism through the dehydration-condensation reaction of \ce{Si(OH)4} forming the Si$-$O$-$Si covalent bond with the coagulated complex phases.
This work highlighted a novel method of ceramic formation based on cement and mineral powder without any heating and pressing action, and further investigations could be expected to verify the improved thermal performance of ceramic-coated concrete wall.

\section*{Acknowledgments}
This research was supported partially from the State Committee of Science and Technology, DPR Korea, under the fundamental research project ``Design of Innovative Functional Materials for Energy and Environmental Application'' (No. 2016-20). The analysis of XRD, EDS and SEM was performed at the Institute of Analysis, Kim Il Sung University, recognized as a certificate institution in DPR Korea. 


\bibliographystyle{elsarticle-num-names}
\bibliography{Reference}

\end{document}